# Predicting Software Defects Through SVM: An Empirical Approach

**Junaid Ali Reshi**
Centre for Computer Science and Technology
Central University of Punjab
Bathinda, Punjab-151001
jreshi14@gmail.com

**Satwinder Singh**
Centre for Computer Science and Technology
Central University of Punjab
Bathinda, Punjab-151001
satwindercse@gmail.com

Abstract- *Software defect prediction is an important aspect of preventive maintenance of a software. Many techniques have been employed to improve software quality through defect prediction. This paper introduces an approach of defect prediction through a machine learning algorithm, support vector machines (SVM), by using the code smells as the factor. Smell prediction model based on support vector machines was used to predict defects in the subsequent releases of the eclipse software. The results signify the role of smells in predicting the defects of a software. The results can further be used as a baseline to investigate further the role of smells in predicting defects.*

## I. INTRODUCTION

Software maintenance is a continuous process that exists throughout the overall lifespan of a software product, software development organizations aim to design the software products that are easy to maintain. There are various factors that help in effective software maintenance by detecting the code anomalies and their symptoms. More specifically, the prediction of faults through metrics, or possible flaws in the software systems through code smells helps in maintainability of a software.

Software metrics, the measure of different parameters of the software code, have also been found to be effective in predicting faults in a software system through the development of various defect prediction models. Software defect prediction is the application of different techniques to predict possible defects in a software. Many methods have been proposed till date to predict the defects in a software in advance, so as to make quality of software better. Machine learning algorithms have been utilized by many researchers to predict defects in a software. Code smell, a symptom of poor design and implementation choice, has also been found to have significant effect on software maintainability. Technically, code smells do not hamper the software system functionality but they tend to increase the risk of system failure in future.

Metric rules and static code analysis have been employed to design various tools for the detection of code smells [1], [2], [3].Some studies have been conducted to determine the side effects of code smells on software maintainability [4], [5], [6], change-proneness [7], [8] and understandability [9]. In their study, Yamashita and Moonen [6] have investigated the relationship between certain inter-smells with some problems during maintenance which also include the introduction of defects to the system.Fontana et al. have experimented 16 different machine learning algorithms with different configurations to experiment techniques for code smell detection [10].

## II. DATA COLLECTION

### A. Source Code selection

The source code selection is an essential component of any analysis. The characteristics of a source code determines the type of results we can infer from them. If we select a source code, big enough to represent an industry sized software, the inferences can hold true for industry software. In addition to this, the platform of the software source code analysed also matters as programming languages differ in principle and an analysis on one platform may not necessarily be applicable on other platform. The source code was selected on the basis of size and platform. Another important aspect of the Eclipse source code is that the source code is open-source and it provides open access to its bug repository and other allied information. This is an important feature as the bug information is vital to know about the presence of Bugs and the related information about the affected classes.

### B. Software Metrics and Code Smells extraction

Software metrics are important depicters of a software. They present vital information about a software vis-à-vis various technical characteristics. The metrics of above-mentioned versions of the software were analysed by iPlasma tool. The source code was also analysed by iPlasma tool for the code smells. The need for analysing the source code for metrics through iPlasma was felt because of the smell prediction model that is to be formed will have some dependencies on such metrics, which were not available through any other extraction software. The source code was also analysed for various smells. The choice of smell was based on the literature survey conducted as well as the options available.

In the dataset creation of the code smells, four class level code smells and three method level code smells were used. The choice of code smells was based on the already available literature and the code smell types available from the tool

The code smells considered (already described) are:

**Table 1.1: Smells considered for extraction from the Source code.**

| Class level Code smells | Method level Code Smells |
| --- | --- |
| Data Class | Brain method |
| God Class (God+Brain) | Feature Envy |
| Refused Bequest | Shotgun Surgery |
| Schizofrenic Class | |

After each code smell was extracted from the source code, these code smells were consolidated to form a single indication of presence of code smells. Each of these method level code smells were resolved to the class level code smells and the metrics considered were that of the class and not the methods, because the investigation was carried at the class level from the onset.

*C. Bug Association*

Out of these source codes, Eclipse 3.2, Eclipse 3.3, Eclipse 3.6, and Eclipse 3.7 were selected on the basis of substantial difference in the build dates and were examined for the presence of bugs. Due to the time limitations and the paucity of bugs in the other products of the Eclipse software, only JDT and PDE components were selected for the resolution of bugs. The bug repository that contains Eclipse bug history and tracking information is Bugzilla.

### III. EXPERIMENTATION

The smell prediction models were created and evaluated for their efficiency on Weka through 10-fold cross validation. The cross-fold validation technique was used to ascertain the relevance of the dataset in predicting the smells. In 10-fold cross-validation, ten different randomisations of the dataset, after dividing the dataset into ten parts (9 for training and one for testing), is used in training and testing of the algorithm. The cross-validation procedure is a standard guaranteeing a stratified sampling of the dataset, reducing the overfitting phenomenon [11] [12] [13], thus providing an efficient way for supervised machine learning.

The defect prediction was performed through the smell prediction models built for each version of the Software.

Different performance measures were obtained and recorded for each run of the model. Each smell prediction model was used to predict defects in the succeeding versions of Eclipse. While performing the testing, the test datasets of Eclipse versions were used down the release versions and not upwards, as chronological feasibility was sought in the study, and rationality demands the same.

Data pre-processing technique, WrapperSubsetEval with Evolutionary search was used to treat the data before creating smell prediction models.

*A. Observations Recorded*

The different parameters used for measuring the efficiency of the models are:

*Precision:*
It is used to measure the degree to which the repeated measurements under unchanged conditions show the same results. It is also called as positive predictive value. Precision refers to the closeness of two or more measurements to each other. Value near one is good. Precision is related to random errors.

$$\text{Precision} = \frac{TP}{TP+FP} \quad (1.1)$$

*Recall:*
It is the measure of the degree of the correctness of the algorithm while classifying a particular class. However, it does not take into account, the wrongly classified instances that are included into the class (False positives). In other words, it is the fraction of positive instances that are classified as positive. In binary classification, it is also called as sensitivity.

$$\text{Recall} = \frac{TP}{TP+FN} \quad (1.2)$$

*F-Measure:*
F-measure is the harmonic mean of precision and recall. Mathematically,

$$Fmeasure = 2 \frac{precision * recall}{precision + recall} \quad (1.3)$$

*ROC (Receiver Operating Characteristics):*
The performance of a binary classifier is also depicted by ROC curve. The graphical representation of ROC curve is the plot between true positive rate (TPR) and false positive rate (FPR) at various threshold values.

### IV. RESULTS AND DISCUSSION

The performance of the SVM based defect smell-defect prediction model is average and doesn't show any exceptional figures. Although the precision of the model lies above 80 percent, yet the recall values, in conjugation lie around 68 percent. This implies that the algorithm has considerable efficiency as the F-measure also tends around 75 percent. Another measure for the goodness of the model is Receiver operator characteristics (ROC) value. These values denote the perceptiveness of the model in discriminating the values. The model shows an average perceptiveness while discriminating the values.

Table 4.6: Performance measures of SVM based Smell-Defect model

| Smell Prediction Model | Defects Predicted in | Precision | Recall | F-Measure | ROC |
|---|---|---|---|---|---|
| Eclipse 3.2 | Eclipse 3.2 | 0.821 | 0.682 | 0.730 | 0.625 |
| | Eclipse 3.3 | 0.816 | 0.695 | 0.737 | 0.641 |
| | Eclipse 3.6 | 0.846 | 0.690 | 0.743 | 0.652 |
| | Eclipse 3.7 | 0.878 | 0.679 | 0.750 | 0.627 |
| Eclipse 3.3 | Eclipse 3.3 | 0.818 | 0.728 | 0.761 | 0.647 |
| | Eclipse 3.6 | 0.845 | 0.721 | 0.765 | 0.651 |
| | Eclipse 3.7 | 0.877 | 0.713 | 0.774 | 0.628 |
| Eclipse 3.6 | Eclipse 3.6 | 0.843 | 0.690 | 0.742 | 0.643 |
| | Eclipse 3.7 | 0.879 | 0.683 | 0.752 | 0.630 |
| Eclipse 3.7 | Eclipse 3.7 | 0.879 | 0.683 | 0.753 | 0.630 |

The area under curve (AUC) values for ROC are indicative of the average performance of the model in determining the defects in software. These curves are plotted for observation

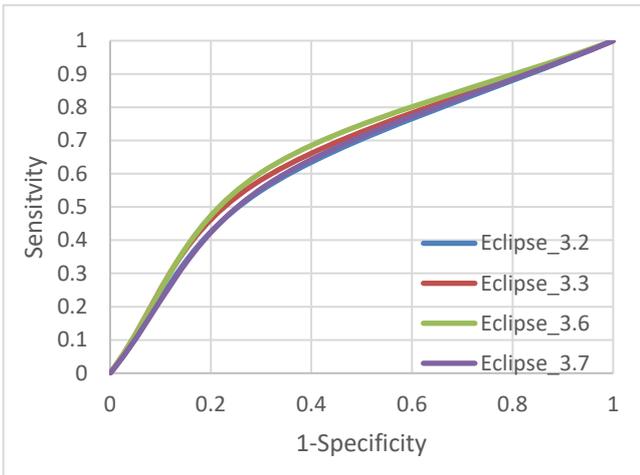

(a)

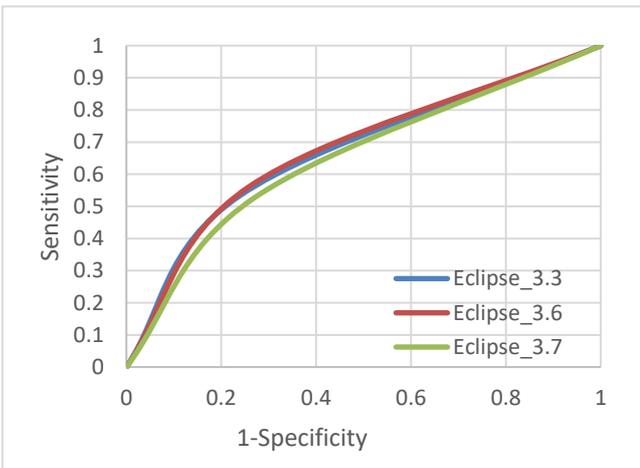

(b)

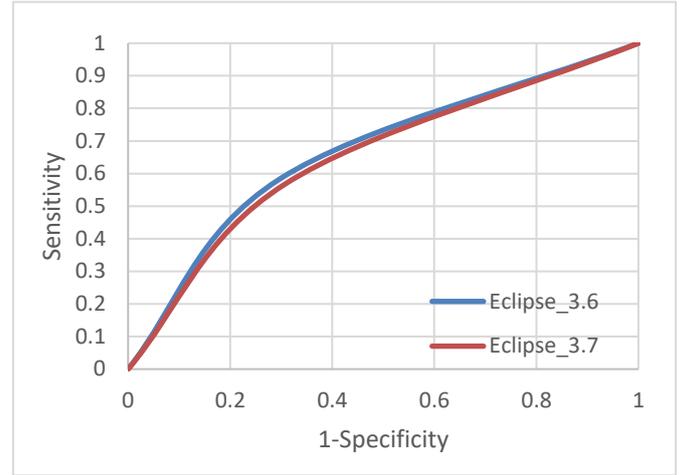

(c)

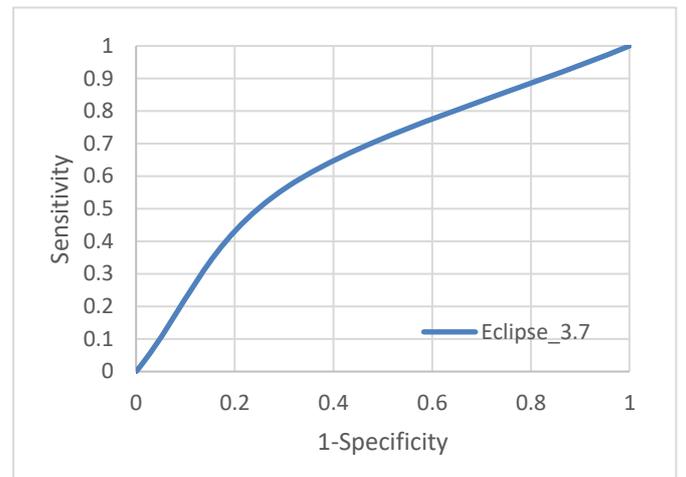

(d)

*Figure 1.11: ROC curves on application of SVM based smell prediction model for prediction of defects in the subsequent versions. (a) Application of Eclipse 3.2 smell prediction model for defect prediction. (b) Application of Eclipse 3.3 smell prediction model for defect prediction. (c) Application of Eclipse 3.6 smell prediction model for defect prediction. (d) Application of Eclipse 3.7 smell prediction model for defect prediction.*

## V. CONCLUSION

This study focusses on the improvement in the process of software maintenance through improving the process of defect prediction. Software smell models, based on the machine learning algorithms, have been used to predict the defects in the software. The results are indicative of the fact that the code smells definitely do have a role in the process of defect prediction. This study can be used a base-line to further explore the relation between code-smells and defects which may further enhance the software maintenance. The work can be extended by encompassing different platforms of industrial software with verified bug-datasets along-with the performance tuning of SVM or by using different machine learning algorithms. In addition, many smell detection techniques can be augmented to present a unbiased version of possible smells which may help us to enhance the smell data. Further, the smells can be dealt with, on individual levels, as well as on the standards already defined for their categorisation and the effectiveness of the models can be studied thereof.